# Long-Term Recurrent Convolutional Network-based Inertia Estimation using Ambient Measurements

Mingjian Tuo  
*Student Member, IEEE*  
Department of Electrical and Computer Engineering  
University of Houston  
Houston, TX, USA  
mtuo@uh.edu

Xingpeng Li  
*Member, IEEE*  
Department of Electrical and Computer Engineering  
University of Houston  
Houston, TX, USA  
xli82@uh.edu

*Abstract*—Conventional synchronous machines are gradually replaced by converter-based renewable resources. As a result, synchronous inertia, an important time-varying quantity, has substantially more impact on modern power systems stability. The increasing integration of renewable energy resources imports different dynamics into traditional power systems; therefore, the estimation of system inertia using mathematical model becomes more difficult. In this paper, we propose a novel learning-assisted inertia estimation model based on long-term recurrent convolutional network (LRCN) that uses system wide frequency and phase voltage measurements. The proposed approach uses a non-intrusive probing signal to perturb the system and collects ambient measurements with phasor measurement units (PMU) to train the proposed LRCN model. Case studies are conducted on the IEEE 24-bus system. Under a signal-to-noise ratio (SNR) of 60dB condition, the proposed LRCN based inertia estimation model achieves an accuracy of 97.56% with a mean squared error (MSE) of 0.0552. Furthermore, with a low SNR of 45dB, the proposed learning-assisted inertia estimation model is still able to achieve a high accuracy of 93.07%.

## I. INTRODUCTION

Inverter-based renewable energy sources (RES) are replacing traditional synchronous generators with the primary goal of carbon dioxide emission reduction and environmental benefits. Due to the increasing grid integration of inverter-based resources such as wind power, solar photovoltaics (PV) and energy storage systems (ESS), system inertia traditionally coming from synchronous generators decreases significantly [1]. System with insufficient synchronous inertia is more likely to suffer high rate of change of frequency (RoCoF) and large frequency excursion during a contingency, resulting in under frequency load shedding (UFLS) as well as tripping of frequency related generator protection devices; the failure of successive units would furthermore cause cascading outages [2].

Inertia estimation can help market design for ancillary services and improve power system reliability through implementation of frequency control ancillary services [3]. Traditionally, system frequency response is analyzed by looking at the collective performance of all generators using a system equivalent model. The system equivalent inertia constant is determined by the number and size of actively connected synchronous units. However, the variability nature of RES imports uncertainties into the system inertial response as well as system inertia constant [4]. Recent study in [5] shows that control schemes can be used to emulate synchronous machine response; such concept introduces techniques like virtual inertia. Additionally, RES such as wind power is interfaced to the grid through converters which electrically decouples the rotor's inertia from the grid, thus RES and other inverter-based sources are traditionally considered passive in terms of inertial response. Therefore, the system inertia constant can only be estimated using ambient wide area measurements.

Traditionally the ability of inertia estimation is mostly dependent on factors like size of disturbance, accuracy of frequency measurement and location of measurement point relative to in-feed loss [6]. Inertia estimation using ambient wide area measurements was proposed in [7]. The method divides the system into a number of subareas and estimates inertia of each subarea separately, but the approximation made in mathematical model introduces more error in the final results. The Electric Reliability Council of Texas (ERCOT) uses a real-time sufficiency monitoring tool to monitor inertia based on the operating plans submitted by the generation resources [8]. Modern power systems are connected to different devices which provide frequency regulation service; meanwhile considering the inertia contribution from demand side, the inertia constant estimation based purely on synchronous generators is inaccurate [9]. Inertia estimation based on mathematical model is also highly dependent on accuracy of ambient measurements from phasor measurement units (PMUs) or equivalent devices, and the estimated value may suffer inaccuracy in various conditions. As RES penetration level increases, the swing equation-based models may no longer represent the system dynamics.

A neural network-based inertia estimation technique is proposed in [10]. The proposed method uses inter-area model information as neural network inputs and estimates the inertia constant as an output of the network. However, this approach only estimates the inertia constant for large systems with only traditional synchronous generation. A convolutional neural network (CNN) based model is proposed in [11], which estimates the system inertia through frequency response and

RoCoF data; only equivalent frequency measurements are considered in this approach, such that results may suffer high errors when non-monotonic frequency deviation occurs.

In this paper, we propose a model-free ambient measurements based machine learning approach to dynamically estimate the system inertia constant. Long-term recurrent convolutional network (LRCN) is used to identify spatial features of the input data and process sequential data. Ambient wide measurements obtained from the PMUs are selected as candidate features for the proposed model. The major contributions of this work are: (a) an LRCN based algorithm is proposed; (b) a wrapper feature selection is used to optimize the feature combination set; (c) ambient measurements under multiple conditions are examined, which improves the estimation accuracy as well as estimator robustness.

The remainder of this paper is organized as follows. In section II, the frequency dynamics of power systems are described. Section III details the proposed inertia estimation algorithm using LRCN. Section IV describes the simulation setup, and the results and analysis are presented in Section V. Section VI presents the concluding remarks and future work.

## II. SYSTEM FREQUENCY DYNAMICS

The inertia constant is a parameter describing the ability of synchronous generator in counteracting the frequency excursion due to power imbalance occurring in power systems. The energy stored in large rotating generator and some industrial motors gives them the tendency to remain rotating. The rotational energy $E_i$ in the rotor of the machine at nominal speed is defined by the following formula:

$$E_i = \frac{1}{2} J_i \omega_i^2 \qquad (1)$$

where $J_i$ is the moment of inertia of the shaft in kg·m²s and $\omega_i$ is the nominal rotational speed. The inertia constant $H_i$ is then given in seconds, which can be expressed as:

$$H_i = \frac{J_i \omega_i^2}{2 S_{B_i}} \qquad (2)$$

where $S_{B_i}$ is the generator rated power in MVA. When multiple generators connected to the power system, dynamics of these generators' rotors are directly coupled with the grid electrical dynamics. Thereby the power system could be represented by a single equivalent model of inertia. The total power system inertia $E_{sys}$ is then considered as the summation of the kinetic energy stored in all dispatched generators synchronized with the power system. It can be shown in the form of either the stored kinetic energy or inertia constants as follows.

$$E_{sys} = \sum_{i=1}^{N} \frac{1}{2} J_i \omega_i^2 = \sum_{i=1}^{N} H_i S_{B_i} \qquad (3)$$

The inertia constant of the power system in seconds is given by the equation below,

$$H_{sys} = \frac{\sum_{i=1}^{N} H_i S_{B_i}}{S_B} \qquad (4)$$

where power $S_B$ is the total rated power of the whole system.

The simplified system equivalent model is based on the extension of one-machine swing equation. For a single machine, the dynamic of its rotor can be described in (5) with $M = 2H$ denoting the normalized inertia constant and $D$ denoting damping constant respectively.

$$\Delta P_m - \Delta P_e = M \frac{d \Delta \omega}{dt} + D \Delta \omega \qquad (5)$$

where $\Delta P_m$ is the total change in mechanical power and $\Delta P_e$ is the total change in electric power from the power system. $d\Delta\omega/dt$ is commonly known as rate of change of frequency (RoCoF).

As most inertia estimation approaches rely on event transient measurement of collective system model following recorded disturbances, studies in [12] have found that the equivalent system model may be affected by inertia heterogeneity and thus cause issues in system operations. Therefore, dynamic model is preferred in modern power system analysis. Using the topological information and system parameters, when multiple generators connected in a bus, as an approximation, equivalent equation (5) can be extended and applied to all buses to describe the oscillatory behavior of each individual bus,

$$m_i \ddot{\theta}_i + d_i \dot{\theta}_i = p_{in,i} - p_{e,i} \qquad (6)$$

where $m_i$ and $d_i$ denote the inertia coefficient and damping ratio for node $i$ respectively, while $p_{in,i}$ and $p_{e,i}$ refer to the power input and electrical power output, respectively. Different from focusing on the collective performance of the power system, the frequency response experienced by each bus could be very distinct. Accordingly, the ambient measurements from the system would provide more information of each subarea and thus improve the accuracy of estimation model.

## III. INERTIA ESTIMATION USING LRCN

### A. System Perturbation using Probing Signal

Probing signal is a method to conduct power system dynamic studies in situations involving system perturbation without affecting system stability [13]. A sample probing signal, fed to the system with an amplitude of $P_E$, and corresponding PMU measurements are shown in Fig. 1.

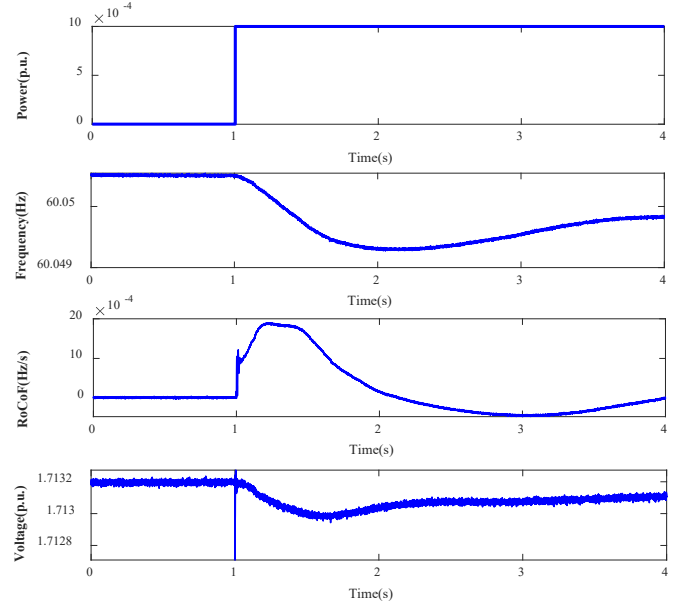

Fig. 1. A sample of probing signal, ambient measurements for $P_E$=0.001 p.u.

With varying system inertia and probing signal amplitude, a number of ambient measurements of $\Delta\omega$, $\Delta\dot{\omega}$ and $v$ can then be collected.

## B. Inertia Estimation using LRCN

Estimation of inertia constant is quite challenging due to the non-linear nature of the power system. Since LRCN leverages the strength of rapid progress in CNN and has the ability to capture the dependencies in a sequence, it has been successfully used in computer vision, image processing, and other fields in signals and time-series analysis [14]. The architecture of proposed LRCN model is displayed in Fig. 2 and can be trained to estimate system inertia from ambient measurements obtained from the PMU.

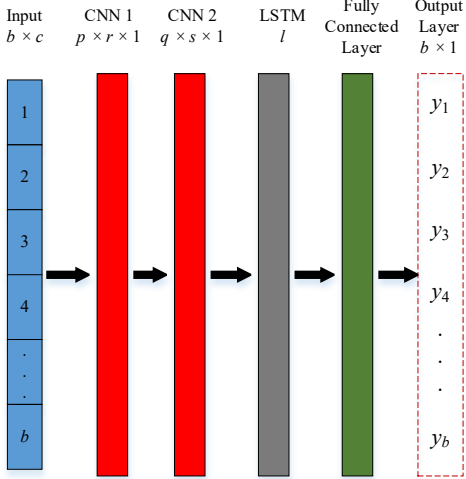

Fig. 2. General architecture of proposed LRCN model.

The proposed LRCN model processes the measurements input with CNN layers first, whose outputs are then fed into long short-term memory (LSTM) recurrent sequence model, and the fully connected layer finally produces inertia constant estimation. The entire dataset is divided into two sets before training: training set (80%) and testing set (20%). The samples in training set are defined in batches which will be propagated through the networks. One epoch of training is completed when all the training samples have been passed forward and backward once. The number of iterations is defined as the number of passes, and each pass uses the same batch size that is the number of samples. At each training iteration, the LRCN model input size is $b \times c$, and the output will be a column vector of size $b$ with inertia estimates for corresponding input in the batch. The dimension of $c$ is determined by the set of features and feature sampling rate.

The mean squared error (MSE) measures the average squared difference between actual and predicted outputs. The goal of training is to minimize MSE via back propagation which will provides best estimator [15]. The fully connected network used in this model includes one flatten layer and two hidden layers. MSE is defined as:

$$MSE = \frac{1}{n}\sum_{i=1}^{n}(y_i - \tilde{y}_i)^2 \quad (7)$$

where $n$ is the total number of training samples, $y_i$ is the actual value of $i^{th}$ output, and $\tilde{y}_i$ is the estimated value corresponding to the $i^{th}$ output. Similarly, the weight update equation via back propagation is expressed as:

$$w_{t+1} = w_t - \alpha \frac{\partial E_{MSE}}{\partial w_t} \quad (8)$$

where $w_t$ is the weight for current iteration, $w_{t+1}$ is the updated weight for next iteration, $\alpha$ is the learning rate, and $E_{MSE}$ is the MSE obtained from expression (7).

## IV. SIMULATION SETUP

### A. Overview

The IEEE 24-bus system [16] was used for the experiment to collect the training data. The system has 24 buses (17 buses with loads), 38 branches, and 38 generators. System inertia $M$ typically ranges from 3s to 8s. Hence, the measurements snapshots were collected for 11 different values of $M$ from 3s to 8s with an increment of 0.5s. Similarly, probing signals with 100 different values of $P_E$ from 0.001 p.u. to 0.01 p.u. with an increment 0.0001 p.u. were used.

The modeling and simulation of the power system, along with data collection, were conducted in MATLAB/Simulink 2019b. The data pre-processing was conducted in both MATLAB and Python. The proposed LRCN model including CNN and LSTM layers was developed in Python using Keras.

### B. Data Preprocessing

The initial data analyzed in this study were acquired from PMUs with a sampling rate of 2880 Hz; for each nodal measurement of $\Delta\omega$ and $\Delta\dot{\omega}$ it gives 2880 data points at a sampling frame of 1s. By using only one second sampling frame for normalization, the real-time applicability of this method is maintained. Similarly, following the same pattern we obtained 2880 data points per second for nodal voltage measurement $v$. Since the training data come from the ambient measurements of all PMUs, without dimension reduction process the original training data would increase the complexity of the model and may also lead to overfitting. Therefore, we first downsample the data of all measurements to 200 Hz similar to [11]. Next, we manually add the additional Gaussian noise signal in the constituent tonic to mimic the noisy measurements. Different signal-to-noise ratios (SNR) are investigated in this paper. Since we analyze the data collected between multiple sessions, the measurements are normalized by employing min/max normalization and thus all input data ranges between [0, 1].

### C. Feature Selection

To find the best time frame of data extraction, different time windows of the ambient measurements are determined: (1) the time frame is first chosen from 0s to 1s following the perturbation, where initial RoCoF is included; (2) the second time frame is from 0.5s to 1.5s after the signal infeed. With $\Delta\omega$ and $\Delta\dot{\omega}$ as basic features combination, the coefficient of determination and validation accuracy are used as evaluating metrics.

In order to determine the optimum set of features for use in the estimation of the system inertia constant, a wrapper feature selection is utilized using (i) the proposed LRCN model as the estimator, (ii) accuracy score with a tolerance of 10% as the evaluation metric, (iii) greedy forward selection as the subset selection policy.

### D. Hyperparameters Selection

With a resampling rate of 200Hz, measurements on each node gives 200 data points at a sampling frame of 1s. In this paper, we first consider the ambient measurements of $\Delta\omega$ and $\Delta\dot{\omega}$ on generator buses and thus we obtain the base vector with dimension of $c = 400 \times 10$. For the convolution layers, the channel number are set $p = 10$ and $q = 20$, and kernels with sizes $r = s = 3$. Rectified linear unit (ReLU) is used as the activation function. The memory unit value $l$ of LSTM layer is set as 32 and the number of neurons in the flatten layer can be manually calculated as $f = 79,960$. The training was operated in batches of 32 data points. An MSE based dynamic learning rate strategy is used for the training. Learning rate schedule is applied in the training process by reducing the learning rate accordingly, the factor by which the learning rate will be reduced is set 0.5 and the patience value is set 10 epochs.

## V. RESULT ANALYSIS

A total of 1,100 samples were collected, the entire dataset is first divided into two subsets: 880 samples (80%) for training and 220 samples (20%) for validation. To leverage the fast-computing abilities of Keras, the machine learning model was trained on NVIDIA Quadro RTX 8000 GPUs.

### A. Time Frame Selection

Measurements $\Delta\omega$ and $\Delta\dot{\omega}$ are selected as training feature combination [11]; training data extracted from two periods are then fed into the proposed LRCN model. Fig. 3 compares the scatter points predicted by the proposed LRCN model with features extracted from 0.5s - 1.5s and 0.0s - 1.0s respectively.

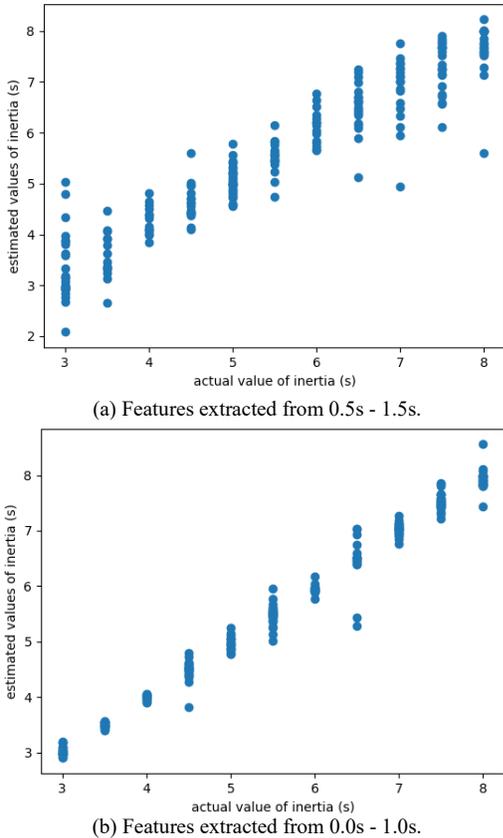

(a) Features extracted from 0.5s - 1.5s.

(b) Features extracted from 0.0s - 1.0s.

Fig. 3. System inertia constant prediction results with the proposed LRCN model using features extracted from different time periods.

The coefficient of determination of the model using features extracted from 0.0s - 1.0s is 0.9625 which outperforms the use of features extracted from 0.5s - 1.5s at 0.7619. As expected, the LRCN model using features extracted from 0.0s - 1.0s has a validation accuracy of 97.56% with a tolerance of 10%, while it is only 75.64% for the use of features extracted from 0.5s - 1.5s.

Thus, we consider that features extracted from the time frame following the disturbance contain prominent inertial response, and accordingly have a positive impact on the overall performance of inertia constant estimation model.

### B. Analyzing the Performance Metrics

Table I compares the performance of LRCN models utilizing different feature combinations. It can be observed that $\Delta\omega$ and $\Delta\dot{\omega}$ is selected as the optimized set of features which outperforms other feature combinations.

Table I
Comparison of different features sets for the proposed LRCN model

| Features Set | $\Delta\omega$ | $\Delta\dot{\omega}$ | $\Delta\omega + \Delta\dot{\omega}$ | $\Delta\omega + \Delta\dot{\omega} + v$ |
|---|---|---|---|---|
| Validation Accuracy | 74.43% | 91.68% | 97.56% | 95.16% |
| Mean Squared Error | 0.2756 | 0.1704 | 0.0552 | 0.1023 |
| Coefficient of Determination | 0.8618 | 0.9234 | 0.9625 | 0.9416 |

Table II
Comparison of different models

| Model | Validation Accuracy | Coefficient of Determination | Mean Squared Error |
|---|---|---|---|
| CNN | 97.56% | 0.9219 | 0.0998 |
| LRCN | 89.59% | 0.9625 | 0.0552 |

The proposed LRCN based approach are compared with benchmark CNN algorithm [11] in Table II. Both algorithms are employed to train the inertia constant estimation model. The results show that the coefficient of determination of LRCN model is 0.9625 which is higher than the coefficient of determination (0.9219) of CNN model. Additionally, the proposed LRCN model has a validation accuracy of 97.56%, with an MSE of 0.0552. For the benchmark CNN model, $\Delta\omega$ and $\Delta\dot{\omega}$ are used as primary features to train the model, resulting in a coefficient of determination of 0.9219, with an MSE of 0.0998. The validation accuracy of CNN model with 10% tolerance is calculated as 89.59%, which is much lower than the proposed LRCN model. The results demonstrate that the optimal set feature is achieved by utilizing wrapper feature selection approach, and thus improves the algorithm performance. Fig. 4 presents the evolution of MSE losses on the training and validation sets over the training process of the proposed LRCN model. MSE decreases as the number of epochs increases.

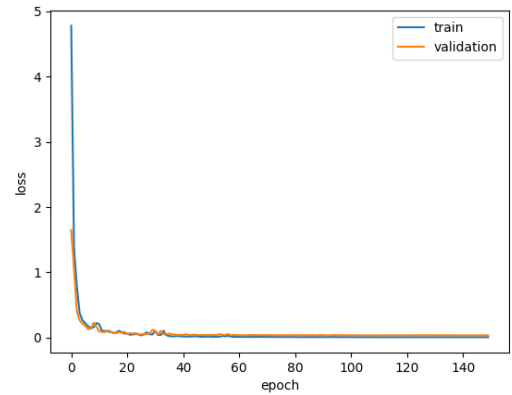

Fig. 4. The learning curve of the proposed LRCN model: MSE losses versus the number of epochs.

### C. Performance with Lower SNR

In addition to the ideal condition, the proposed algorithm is compared with the benchmark CNN model under high noise conditions. Study in [17] has shown that a SNR of 45dB is considered a good approximation of noise power under realistic condition. Fig. 5 shows the inertia constant estimates from CNN model with $\Delta\omega$ and $\Delta\dot{\omega}$ as training features. After adding additional Gaussian noise signal with a SNR of 45dB to the ambient measurements, the overall MSE of traditional CNN model increases from 0.0998 to 0.1816, while the coefficient of determination reduces from 0.9219 to 0.8852. Understandably, a significant reduction in validation accuracy can be observed, which drops from 89.59% to 76.36%.

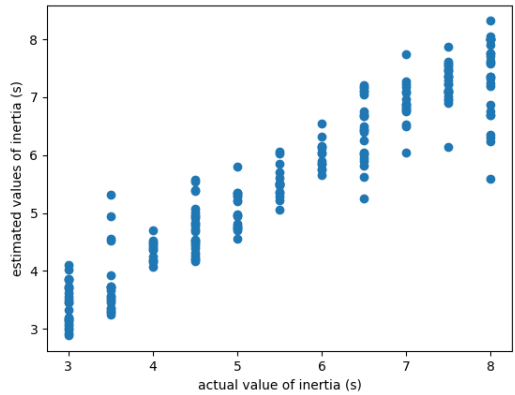

Fig. 5. Prediction results of the benchmark CNN model with SNR at 45dB.

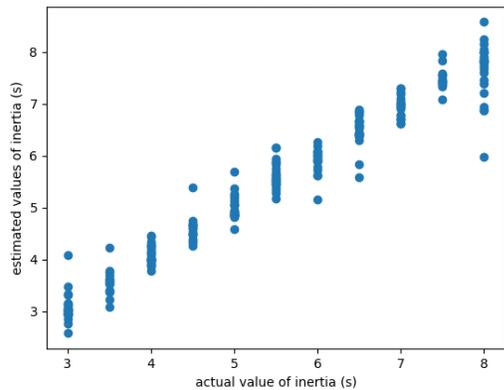

Fig. 6. Prediction results of the proposed LRCN model with SNR at 45dB.

The prediction results of the proposed LRCN model are presented in Fig. 6. The method described in this research uses a wrapper feature selection process and then, the measurements of $\Delta\omega$, $\Delta\dot{\omega}$ and $v$ are selected as the optimal set of features used for inertia constant estimation. The results show that the proposed LRCN model has a validation accuracy of 93.07%, with an MSE of 0.1545. In summary, (i) under a low noise condition with SNR of 60dB, measurements of $\Delta\omega$ and $\Delta\dot{\omega}$ are the optimal set of features suitable for inertia estimation; (ii) under high noise condition with SNR of 45dB, the performance of benchmark model decreases significantly, while the proposed LRCN model based on optimal features combination of $\Delta\omega$, $\Delta\dot{\omega}$ and $v$ shows higher robustness and better performance.

## VI. CONCLUSIONS

Neural networks have been applied to inertia estimation as extensive amounts of data can be obtained from power system digital equipment and advanced measuring infrastructures such as PMU. In this paper, an LRCN based learning algorithm is proposed to estimate system inertia constant. System wide ambient measurements are used as candidate features for model training, and a wrapper feature selection is also used to optimize the feature combination. Results demonstrate that the proposed LRCN algorithm has a better performance than the benchmark CNN model in the literature. The proposed algorithm also shows high robustness under conditions with higher noise. Considering that the IEEE 24-bus system model used in this research has a mix generation of both synchronous generators and inverter-based resources, the proposed approach can also be applied to estimate inertia constant in realistic conditions.